\documentclass[twocolumn,showpacs,preprintnumbers,amsmath,amssymb,floatfix,prl]{revtex4}

\usepackage{epsfig}
\begin{document}

\preprint{BNL-NT-08/8}

\title{Global Analysis of Helicity Parton Densities and Their Uncertainties}
\author{Daniel de Florian}  
% \email{deflo@df.uba.ar}
%
\author{Rodolfo Sassot} 
% \email{sassot@df.uba.ar}
\affiliation{Departamento de Fisica, Universidad de Buenos Aires, 
Ciudad Universitaria, Pabellon 1 (1428) Buenos Aires, Argentina}
\author{Marco Stratmann}  
% \email{marco@ribf.riken.jp}
\affiliation{Radiation Laboratory, RIKEN, 2-1 Hirosawa, Wako, Saitama 351-0198, Japan}
\author{Werner Vogelsang} 
%\email{vogelsan@quark.phy.bnl.gov}
\affiliation{Physics Department, Brookhaven National Laboratory, Upton, NY 11973}

\begin{abstract}
We present a new analysis of the helicity parton distributions of the
nucleon. The analysis takes into account the available data from
inclusive and semi-inclusive polarized deep inelastic scattering, 
as well as from polarized pp scattering at RHIC. For the first
time, all theoretical calculations are performed fully at next-to-leading 
order (NLO) of perturbative QCD, using a method that allows to 
incorporate the NLO corrections in a very fast and efficient way
in the analysis. We find evidence for a rather small gluon polarization 
in the nucleon, over a limited region of momentum fraction, and for
interesting flavor patterns in the polarized sea.

\end{abstract}


\maketitle

%%%%%%%%%%%%%%%%%%%%%%%%%%%%%%%%%%%%%%%%%%%%%%%%%%%%%%%%%%%%%%%%%%%%%%%%
%%%%%%%%%%%%%%%%%%%%%%%%%%%%%%%%%%%%%%%%%%%%%%%%%%%%%%%%%%%%%%%%%%%%%%%%

{\it Introduction.---}The exploration of the inner structure of the 
nucleon is of fundamental
importance in Nuclear and Particle Physics. Of particular interest is
the spin structure of the nucleon, which addresses questions such as
how the nucleon spin is composed of the spins and orbital angular momenta 
of the quark and gluons inside the nucleon. In deep inelastic scattering
(DIS) of leptons off polarized nucleons it was found that surprisingly little 
of the proton spin is carried by the quark and antiquark 
spins~\cite{Bass:2004xa}. This has triggered much theoretical
progress, and led to new experiments dedicated to unraveling
the proton spin structure. Among them, experiments in polarized
proton-proton collisions at the BNL Relativistic Heavy Ion Collider, 
RHIC, have recently opened a new stage in this quest. 
 
The structure of a nucleon in a helicity eigenstate is foremost 
described by the (anti)quark and gluon helicity parton distribution 
functions (PDFs), defined by
\begin{equation}
\label{eq:pdf}
\Delta f_j(x,Q^2) \equiv f^+_j(x,Q^2) -
                       f^-_j(x,Q^2).
\end{equation}
Here, $f^+_j(x,Q^2)$ [$f^-_j(x,Q^2)$] denotes the distribution
of a parton of type $j$ with positive [negative] helicity in 
a nucleon with positive helicity, having light-cone momentum
fraction $x$ of the nucleon momentum and being probed at a 
hard scale $Q$. The integral 
$\Delta f_j^1(Q^2)\equiv\int_0^1 \Delta f_j(x,Q^2) dx$  
measures the spin contribution of parton $j$ to the proton spin, which is 
one reason why there are world-wide efforts to extract the $\Delta f_j(x,Q^2)$ 
from experimental data. 

The non-perturbative but universal $\Delta f_j$ are accessible in
measurements of double-spin asymmetries,
\begin{equation}
\label{eq:asydef}
A_{{\mathrm{LL}}}\equiv\frac{d\Delta \sigma}{d\sigma}\equiv\frac{ 
d\sigma^{++} - d\sigma^{+-}}{d\sigma^{++} + d\sigma^{+-}} ,
\end{equation}
for processes characterized by large momentum transfer
and helicity settings $\pm$.
Taking high transverse momentum ($p_T$) reactions in polarized
pp scattering as an example, the cross section at hadron-level
schematically reads up to corrections suppressed by inverse powers of
$p_T$:
\begin{eqnarray} 
\label{eq1old}
d\Delta\sigma&=&\sum_{ab}
\int dx_a \int dx_b \, \Delta f_a(x_a,Q^2) \Delta f_b(x_b,Q^2) 
\nonumber \\[1mm]
&&\times \, d\Delta \hat{\sigma}_{ab}(x_a,x_b,p_T,\alpha_s(Q^2),p_T/Q).
\end{eqnarray}
The sum runs over all initial partons $a,b$, with
$d\Delta \hat{\sigma}_{ab}$ the corresponding partonic 
cross sections, defined in analogy with Eq.~(\ref{eq:asydef}).
We note that depending
on the experimental observable, also an additional fragmentation function 
may occur in Eq.~(\ref{eq1old}).
An equation similar to (\ref{eq1old}) holds for the unpolarized cross section
$d\sigma$.
The $d\Delta \hat{\sigma}_{ab}$
depend only on scales of the order of the hard scale $p_T$ and are hence
amenable to QCD perturbation theory.
A consistent NLO analysis of (\ref{eq1old}) requires use of 
NLO partonic cross sections and scale evolution for the PDFs.

In DIS, an expression analogous to the one in Eq.~(\ref{eq1old}) holds,
except that there is only one parton distribution. Efforts over the past
three decades have produced extensive data sets for polarized 
DIS \cite{ref:disdata}. Results from semi-inclusive 
DIS (SIDIS) \cite{ref:disdata,Airapetian:2008qf}, 
$lN\to lhX$, with $h$ an identified hadron in the final state, have the 
promise to put individual constraints on the various quark flavor distributions 
in the nucleon. 
Recently, the first precise (in part still 
preliminary) $A_{{\mathrm{LL}}}$ measurements from RHIC have emerged
\cite{ref:rhicdata}, which are 
expected to put significant constraints on the helicity gluon 
distribution, $\Delta g(x,Q^2)$, along with results from
lepton nucleon scattering~\cite{ref:otherdeltag}.

This paper presents the first ``global'' NLO analysis of the data from 
DIS, SIDIS, and RHIC in terms of the helicity PDFs. 
While there have been quite a few NLO analyses of the polarized DIS 
data in the 
past (see, e.g., Refs.~\cite{ref:grsv,ref:other,ref:dns}),
some of which \cite{ref:dns} also take into account information 
from SIDIS, the full inclusion of RHIC data in the NLO analysis is a 
new feature. The fact that the cross section in the pp case, 
Eq.~(\ref{eq1old}), is bilinear in the PDFs, its 
more complicated kinematic structure, and the overall high complexity of 
NLO partonic cross sections, present significant
technical challenges in this endeavor. Based on a technique presented in 
\cite{ref:mellin}, we have now developed and commissioned 
a systematic procedure for performing full NLO 
analyses also when using pp scattering data.

%%%%%%%%%%%%%%%%%%%%%%%%%%%%%%%%%%%%%%%%%%%%%%%%%%%%%%%%%%%%%%%%%%%%%%%%%%
%%%%%%%%%%%%%%%%%%%%%%%%%%%%%%%%%%%%%%%%%%%%%%%%%%%%%%%%%%%%%%%%%%%%%%%%%%
{\it Global analysis and Mellin technique.---}The idea behind a global
analysis is to extract the universal PDFs entering factorized cross sections such as
Eq.~(\ref{eq1old}) by optimizing the agreement between the
measured spin asymmetries for DIS, SIDIS, and pp scattering, 
relative to the accuracy of the data, and corresponding
theoretical calculations, through variation of the 
shapes of the polarized PDFs.
To be specific, 
we choose an initial scale for the evolution of $Q_0=1\,
\mathrm{GeV}$ and assume the helicity PDFs to have the
following flexible functional forms:
\begin{equation}
\label{eq:pdf-input}
x\Delta f_j(x,Q^2_0) = N_j x^{\alpha_j} (1-x)^{\beta_j} 
(1+\gamma_j \sqrt{x}+\eta_j x)  ,
\end{equation}
with free parameters $N_j,\alpha_j,\beta_j,\gamma_j,\eta_j$.
The strategy is, then, to evolve the distributions at NLO to the scales 
relevant to the various data points, to use the evolved distributions 
to calculate the NLO theoretical spin asymmetries at the kinematics 
of each data point, and to construct a $\chi^2$ function
testing the goodness of fit.
The optimal parameters are then found by minimizing $\chi^2$. 

The main technical challenge in our analysis 
lies in the inclusion of the RHIC $A_{\mathrm{LL}}$ data.
The $\chi^2$ minimization
procedure easily requires of the order of $10^5$ or more evaluations
of the rather complex NLO pp cross sections \cite{ref:ppxsec} for each
data point, so that the computer time needed for
a minimization directly on the basis of Eq.~(\ref{eq1old}) becomes excessive. 
In Ref.~\cite{ref:mellin}, we devised a method that allows 
to overcome this problem by working in Mellin-moment space.
This is the approach that we will also pursue here.

In short, it amounts to expressing the PDFs in Eq.~(\ref{eq1old})
by their Mellin inverse transformations to find,
\begin{eqnarray} 
\label{eq1mom}
d\Delta\sigma&=&-\frac{1}{4\pi^2} \sum_{ab} 
\int_{{\cal C}_n}dn \int_{{\cal C}_m} dm\,
\Delta f_a^n(Q^2)\, \Delta f_b^m(Q^2)\nonumber \\ 
&\times& \, \int dx_a \int dx_b \, x_a^{-n}
\,x_b^{-m}   d\Delta \hat{\sigma}_{ab}(x_a,x_b,\ldots) ,
\end{eqnarray}
where ${\mathcal C}_{n,m}$ denote appropriate integration contours in 
complex $n,m$-space, see \cite{ref:mellin} for details.
Here, the crucial point is that the information on the PDFs, 
contained in the Mellin moments $\Delta f_a^n$ and $\Delta f_b^m$,
defined as usual by
$
\Delta f_j^n(Q^2)\;\equiv\;\int_0^1 dx \;x^{n-1}\,\Delta f_j(x,Q^2),
$
has been separated from the numerically tedious and time consuming part
involving the 
$d\Delta\hat{\sigma}_{ab}$ in the second row of Eq.~(\ref{eq1mom}).
As the latter do not depend on the PDFs, their values can be 
computed prior to the fitting procedure, on a suitable array of 
moments $n$ and $m$. 
Once this has been done, the remaining inverse Mellin integrals in 
(\ref{eq1mom}) can be performed extremely fast. 
In fact, the calculation of $d\Delta \sigma$ becomes well
over two orders of magnitude faster than for the ``direct'' method of
computing it via Eq.~(\ref{eq1old}). This brings one to the kinds of 
computational speeds needed for a full inclusion of pp scattering data 
in a NLO fitting analysis without having to resort to approximations.
We note that the calculation of the 
arrays just mentioned is a major computational challenge; however, 
we found that this procedure can be made very efficient and fast 
by using adaptive Monte-Carlo sampling techniques.

Another crucial issue to be addressed in a global analysis is the estimate 
of the uncertainties in the extraction of the various 
$\Delta f_j$, associated with either experimental or theoretical
uncertainties. 
%Different strategies have been conceived and explored.
We pursue here an approach based on the use of ``Lagrange multipliers'' (LM)
\cite{ref:dns,ref:lagrange}. It has the advantage that no assumptions regarding,
e.g., a quadratic behavior of the $\chi^2$ function around the 
minimum need to be made. Rather, one investigates how $\chi^2$ varies 
as a function of a particular observable or variable of interest.
The uncertainty range is then defined by the region for which the increase
$\Delta\chi^2$ in $\chi^2$ above its lowest value is tolerable.
In this way, one finds the largest possible range for predictions of a certain
physical quantity that is consistent with a given $\Delta \chi^2$. 
Use of the LM method requires performing a 
huge number of fits, for which the efficiency of our Mellin method is crucial.

%%%%%%%%%%%%%%%%%%%%%%%%%%%%%%%%%%%%%%%%%%%%%%%%%%%%%
%%%%%%%%%%%%%%%%%%%%%%%%%%%%%%%%%%%%%%%%%%%%%%%%%%%%%
%
{\it Results of global analysis.---}We
now present results for our NLO global analysis of DIS, SIDIS,
and (in part preliminary) RHIC data. The data sets we take into account 
\cite{ref:disdata,ref:rhicdata} are listed in Tab.~\ref{tab:chi2table}
together with their respective $\chi^2$ values \cite{ref:footnote1}.
Notice that we use 
new sets of fragmentation functions (FFs) from Ref.~\cite{ref:dss},
which are consistent with all relevant unpolarized SIDIS and RHIC data 
entering the spin asymmetries analyzed here. Uncertainty estimates 
of the FFs \cite{ref:dss} are propagated in the computation of 
$A_{\mathrm{LL}}$ and included in $\chi^2$ as a theoretical error.
We use the strong coupling $\alpha_s$ and unpolarized PDFs of Ref.~\cite{ref:mrst}.
Other sets \cite{ref:cteq} give very similar results.

\begin{figure}[th!]
\begin{center}
%\vspace*{-0.5cm}
\epsfig{figure=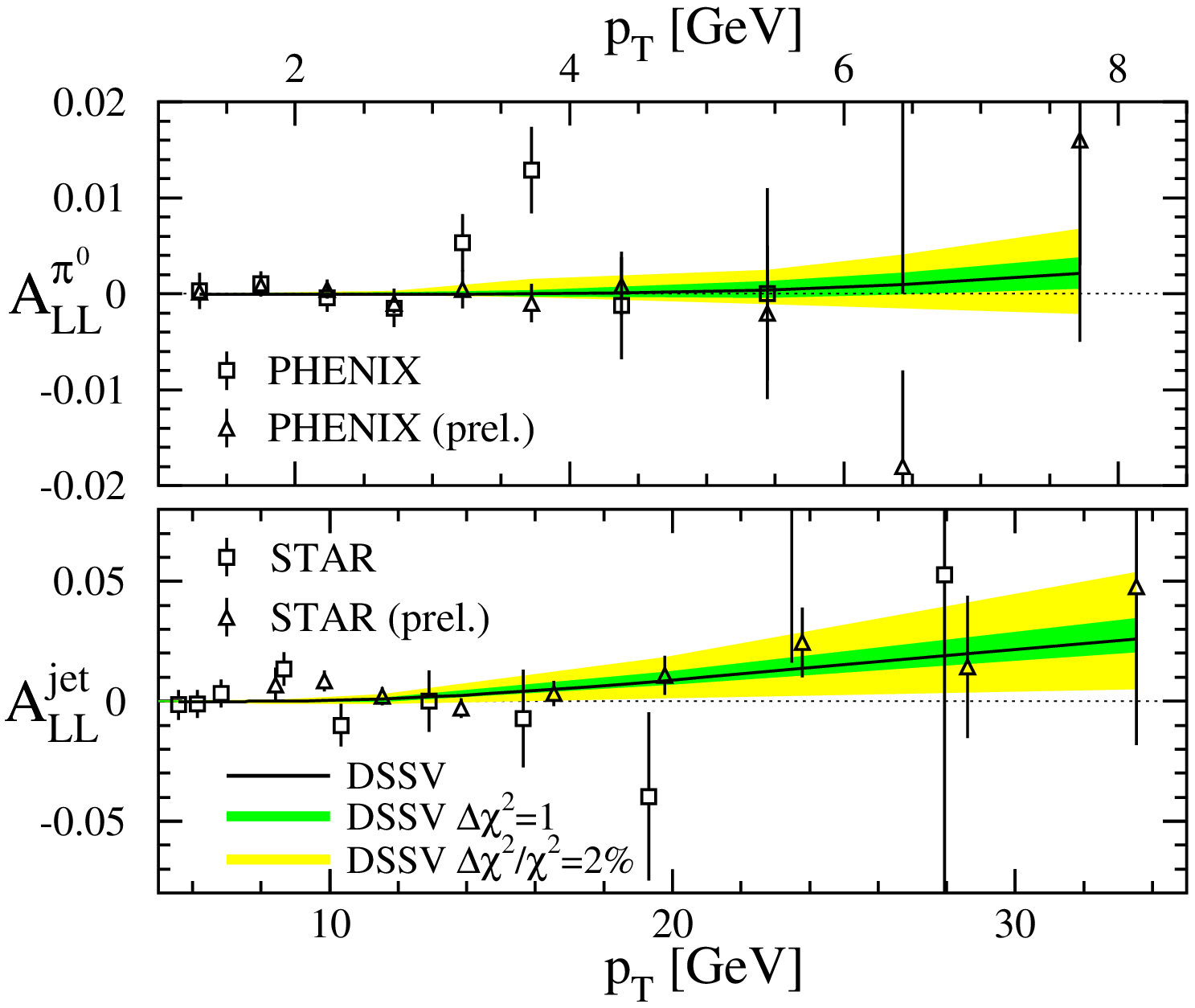,width=0.495\textwidth}
\end{center}
\vspace*{-0.7cm}
\caption{Comparison of RHIC data \cite{ref:rhicdata} and our fit.
The shaded bands correspond to $\Delta\chi^2 = 1$ and 
$\Delta\chi^2/\chi^2 = 2\%$ (see text).
\label{fig:rhic}}
%\vspace*{-0.5cm}
%\end{figure}
%
%%%%%%%%%%%%%%%%%
% FIGURE 2
%%%%%%%%%%%%%%%%%
%\begin{figure}[th!]
\vspace*{-0.3cm}
\begin{center}
%\vspace*{-0.9cm}
\epsfig{figure=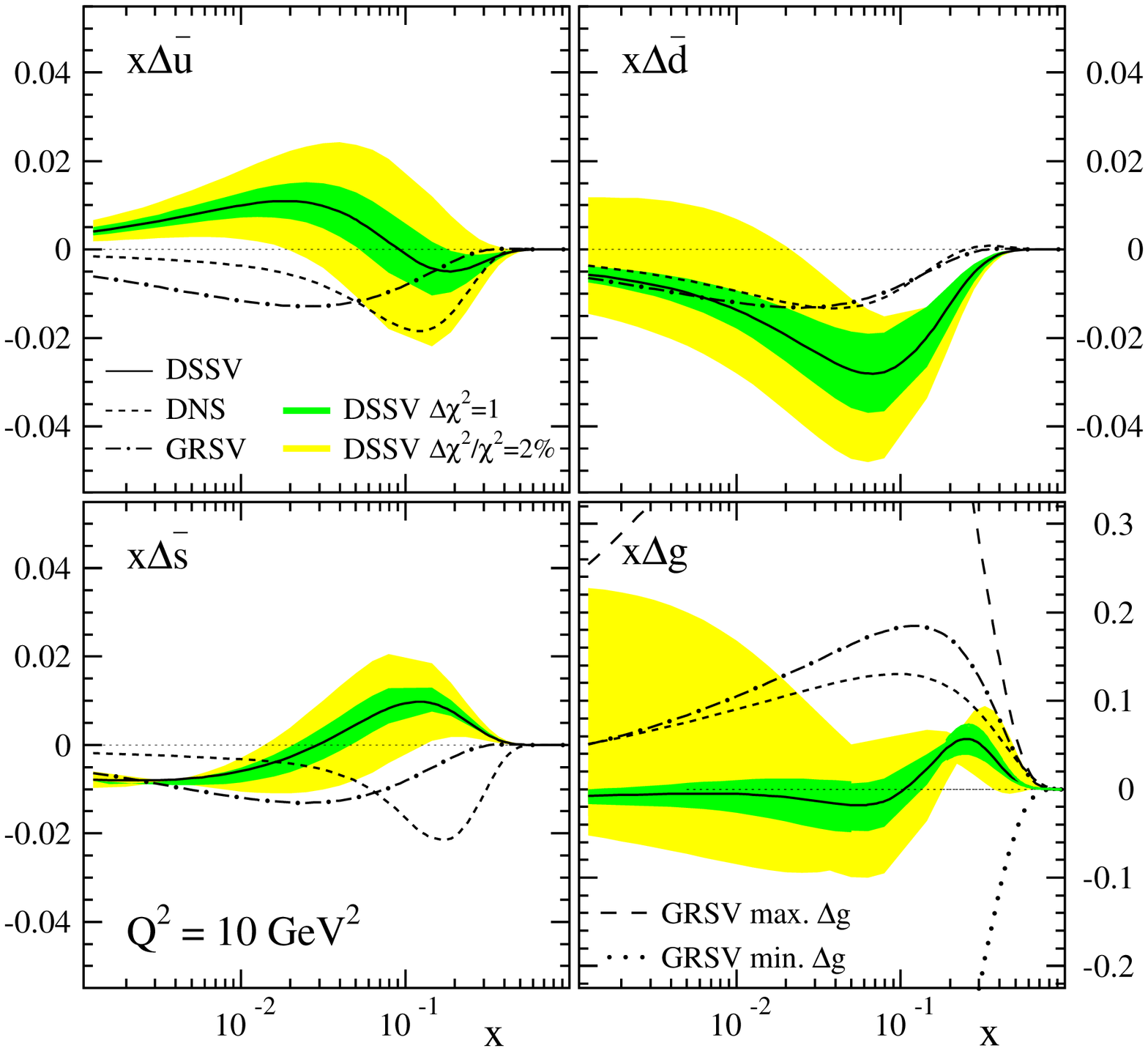,width=0.48\textwidth}
\end{center}
\vspace*{-0.7cm}
\caption{Our polarized sea and gluon densities
compared to previous fits \cite{ref:grsv,ref:dns}.
The shaded bands correspond to alternative fits with $\Delta\chi^2 = 1$ and 
$\Delta\chi^2/\chi^2 = 2\%$ (see text).
\label{fig:uncband}}
%\vspace*{-0.5cm}
\end{figure}

Rather than imposing the standard SU(2) and SU(3) symmetry 
constraints on the first moments of the quark and antiquark distributions,
we allow for deviations: 
\begin{eqnarray}
\label{eq:su2}
\Delta {\cal U} - \Delta {\cal D}= 
(F+D)[1+\varepsilon_{{\mathrm{SU}}(2)}], \\
\label{eq:su3}
\Delta {\cal U} + \Delta {\cal D}-2 \Delta {\cal S} =
(3F-D) [1+\varepsilon_{{\mathrm{SU}}(3)}],
\end{eqnarray}
where $\Delta {\cal F}\equiv\left[ \Delta f^1_j+\Delta \bar{f}^1_j\right] (Q_0^2)$,
$F+D=1.269\pm0.003$, $3F-D=0.586\pm0.031$ \cite{ref:disdata}, 
and $\varepsilon_{{\mathrm{SU}}(2,3)}$ are free parameters.
%
%
%%%%%%%%%%%%%%%%%%%
% TABLE I
%%%%%%%%%%%%%%%%%%
\begin{table}[th!]
\caption{\label{tab:chi2table}Data used in 
our 
analysis 
\cite{ref:disdata,ref:rhicdata}, 
the individual $\chi^2$ values, 
and the total $\chi^2$ of the fit. 
We employ cuts of $Q,\,p_T>1\,\mathrm{GeV}$ for the DIS, SIDIS, 
and RHIC high-$p_T$ data.}
\begin{ruledtabular}
\begin{tabular}{lccc}
experiment& data & data points & $\chi^2$ \\
          & type & fitted      &          \\\hline
%
%DIS
%
EMC, SMC  & DIS     &    34  &  25.7  \\
COMPASS   & DIS     &    15  &  8.1  \\
E142, E143, E154, E155 & DIS     &   123  &  109.9  \\
HERMES    & DIS     &    39  &  33.6  \\
HALL-A    & DIS     &     3  &  0.2  \\
CLAS      & DIS     &    20  &  8.5  \\ \hline 
%
%SIDIS
%
SMC    & SIDIS, $h^{\pm}$   &  48  & 50.7    \\
HERMES & SIDIS, $h^{\pm}$   &  54  & 38.8    \\
       & SIDIS, $\pi^{\pm}$ &  36  & 43.4    \\
       & SIDIS, $K^{\pm}$   &  27  & 15.4    \\
COMPASS & SIDIS, $h^{\pm}$  &  24  & 18.2    \\ \hline
%
%RHIC
%
PHENIX (in part prel.)  &         $200\,\mathrm{GeV}$ pp, $\pi^0$ &  20 & 21.3  \\
PHENIX (prel.)        &   $62\,\mathrm{GeV}$ pp, $\pi^0$  &   5 & 3.1   \\
STAR (in part prel.)     & $200\,\mathrm{GeV}$ pp, jet     &  19 & 15.7  \\ \hline
%
%SUM
%
{\bf TOTAL:} & & 467 & 392.6  \\
\end{tabular}
\end{ruledtabular}
\end{table}
In total we have fitted 26 parameters \cite{ref:webpage},
setting $\gamma_{\bar{u},\bar{d},\bar{s},g}=0$ in Eq.~(\ref{eq:pdf-input}).
Positivity relative to the unpolarized PDFs of Ref.~\cite{ref:mrst}
is enforced at $Q_0$.
In Fig.~\ref{fig:rhic} we compare the results of our fit 
using $Q=p_T$ to RHIC data from 
polarized pp collisions at 200 GeV \cite{ref:rhicdata}, included for the first time 
in a NLO global fit. 
The bands are obtained with the LM method applied to each data point and
correspond to the maximum variations for $A_{\mathrm{LL}}$ 
computed with alternative fits consistent with an increase of 
$\Delta\chi^2 = 1$ or $\Delta\chi^2/\chi^2 = 2\%$ in the total $\chi^2$ of the fit.

\begin{figure}[th!]
\begin{center}
\vspace*{-0.3cm}
\epsfig{figure=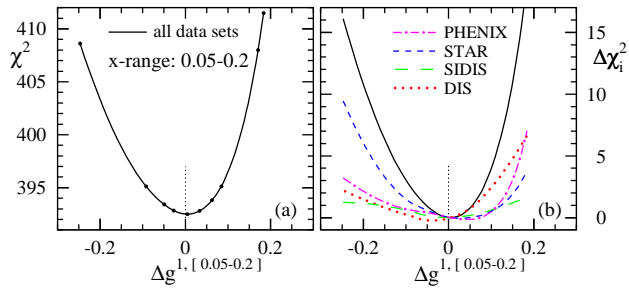,width=0.495\textwidth}
\end{center}
\vspace*{-0.5cm}
\caption{The $\chi^2$ profile (a) and partial contributions
$\Delta \chi^2_i$ (b) of the data sets
for variations of  $\Delta g^{1,[0.05-0.2]}$  at $Q^2=10\,\mathrm{GeV}^2$. 
\label{fig:chi2plot}}
%\vspace*{-0.5cm}
\end{figure}

Our newly obtained antiquark and gluon PDFs 
are shown in Fig.~\ref{fig:uncband} and compared to previous analyses
\cite{ref:dns,ref:grsv}. For brevity, the total $\Delta u+\Delta\bar{u}$
and $\Delta d + \Delta \bar{d}$ densities are not shown 
as they are very close to those in all other fits \cite{ref:grsv,ref:other,ref:dns}.
Here, the bands correspond to 
fits which maximize the variations of the truncated first moments, 
\begin{equation}
\Delta f_j^{1,[x_{\min}-x_{\max}]}(Q^2) \equiv \int_{x_{\min}}^{x_{\max}} \Delta f_j(x,Q^2) dx,
\label{eq:momdef}
\end{equation}
at $Q^2=10\,\mathrm{GeV}^2$ and for $[0.001-1]$.
As in Ref.~\cite{ref:dns} they can be taken as faithful estimates of the 
typical uncertainties for the antiquark densities.
For the elusive polarized gluon distribution, however, we perform a 
more detailed estimate, now discriminating three regions in $x$: 
$[0.001-0.05]$, $[0.05-0.2]$ (roughly corresponding to the range 
probed by RHIC data), and $[0.2-1.0]$. 
Within each region, we scan again for alternative fits that maximize 
the variations of the truncated moments $\Delta g^{1,[x_{\min}-x_{\max}]}$, sharing 
evenly to $\Delta\chi^2$. 
In this way we can produce a larger variety of 
fits than for a single $[0.001-1]$ moment, and, therefore, 
a more conservative estimate. 
Such a procedure is not necessary for antiquarks whose $x$-shape
is already much better determined by DIS and SIDIS data.
%
%%%%%%%%%%%%%%%%%
% FIGURE 1
%%%%%%%%%%%%%%%%%
One can first of all see in Fig.~\ref{fig:uncband} that 
$\Delta g(x,Q^2)$ comes out rather small, 
even when compared to fits with a ``moderate'' gluon 
polarization \cite{ref:grsv,ref:dns}, with a possible node in the distribution. 
This is driven mainly by the RHIC data, which put a strong constraint 
on the size of $\Delta g$ for $0.05\lesssim x \lesssim 0.2$ 
but cannot determine its sign as they mainly probe $\Delta g$ squared.
To explore this further, Fig.~\ref{fig:chi2plot} shows the $\chi^2$ profile 
and partial contributions $\Delta \chi^2_i$ of the individual data sets
for variations of $\Delta g^{1,[0.05-0.2]}$.
A nice synergy of the different data sets is found.
A small $\Delta g$ at $x\simeq 0.2$ is also consistent with data 
from lepton-nucleon scattering \cite{ref:otherdeltag}, which still lack 
a proper NLO description. The small $x$ region remains still largely unconstrained,
making statements about $\Delta g^1$ not yet possible.

%%%%%%%%%%%%%%%%%
% FIGURE 3
%%%%%%%%%%%%%%%%%
%
%
%%%%%%%%%%%%%%%%%
% TABLE II
%%%%%%%%%%%%%%%%%
%
\begin{table}[thb!]
\caption{\label{tab:1stmoments} First moments $\Delta f_j^{1,[x_{\min}-1]}$ at $Q^2=10\,\mathrm{GeV^2}$.}
\begin{ruledtabular}
\begin{tabular}{cccc}
& $x_{\min}=0$ & \multicolumn{2}{c}{$x_{\min}=0.001$} \\
& best fit &$ \Delta \chi^2=1$& $ \Delta \chi^2/\chi^2=2\%$ \\ \hline
$\Delta u + \Delta\bar{u}$  & 0.813  &  0.793 $^{+0.011}_{-0.012}$ & 0.793 $^{+0.028}_{-0.034}$ \\
$\Delta d + \Delta\bar{d}$  & -0.458 & -0.416 $^{+0.011}_{-0.009}$ &-0.416 $^{+0.035}_{-0.025}$ \\
$\Delta\bar{u}$             & 0.036  &  0.028 $^{+0.021}_{-0.020}$ & 0.028 $^{+0.059}_{-0.059}$  \\
 $\Delta\bar{d}$            & -0.115 & -0.089 $^{+0.029}_{-0.029}$ &-0.089 $^{+0.090}_{-0.080}$ \\
$\Delta\bar{s}$             & -0.057 & -0.006 $^{+0.010}_{-0.012}$ &-0.006 $^{+0.028}_{-0.031}$ \\
$\Delta g$                  & -0.084 &  0.013 $^{+0.106}_{-0.120}$ & 0.013 $^{+0.702}_{-0.314}$  \\ 
$\Delta\Sigma$              & 0.242  &  0.366 $^{+0.015}_{-0.018}$ & 0.366 $^{+0.042}_{-0.062}$ \\
\end{tabular}
\end{ruledtabular}
\end{table}
We also find that the SIDIS data give rise to a robust pattern for the 
sea polarizations, clearly deviating from SU(3) symmetry,
which awaits further clarification from the upcoming $W$ boson program at RHIC.
A particularly interesting result emerges for the polarized strange quark distribution: 
a fit that excludes the SIDIS data prefers a negative $\Delta s$, but with
SIDIS data included, $\Delta s$ is forced to be positive for $x\gtrsim 0.02$, 
in agreement with a recent LO analysis in \cite{Airapetian:2008qf}. 
From the fit we find breaking parameters $\varepsilon_{{\mathrm{SU}}(2,3)}$
in (\ref{eq:su2}), (\ref{eq:su3}) 
very close to zero, so that the first moment of $\Delta s$ must be negative. 
Therefore, in the full fit, $\Delta s$ turns negative at small $x$, gaining most
of its area there. 
This is also visible in Tab.~\ref{tab:1stmoments},
where we show the full and truncated first moments of our PDFs.
The behavior of $\Delta s$ also leaves its imprint
on the quark singlet, $\Delta \Sigma$. 
Notice that below $x\simeq 0.001$, where no data are available, 
the contributions to the full moments are determined by extrapolation 
of the distributions rather than constrained by the fit.

{\it Conclusions.---}We have presented the first 
global NLO QCD analysis of DIS, SIDIS, and preliminary RHIC data 
in terms of the helicity parton distributions. A technique
based on the use of Mellin moments of the PDFs allows
to efficiently incorporate the data from pp scattering at RHIC in a full 
and consistent NLO analysis. We have found that the RHIC data set 
significant constraints on the gluon helicity distribution, providing
evidence that $\Delta g(x,Q^2)$ is small in the accessible range of
momentum fraction. We also found that the SIDIS data clearly point to
a mostly positive $\Delta \bar{u}$ and a negative $\Delta 
\bar{d}$. The strange quark distribution $\Delta s$ comes out
negative at $x\lesssim 0.02$ and positive at higher $x$, 
even though here the systematic uncertainties inherent in SIDIS
are arguably largest. 

While our study should and will be improved on a number of aspects,
in particular related to the inclusion of theoretical uncertainties and
the treatment of experimental ones, we believe that it opens the door 
to finally obtaining a better and more reliable picture of the spin 
structure of the nucleon. In particular, it will help RHIC to realize 
its full potential, as hopefully more and more precise data will emerge
over the next few years. We finally note that use of our fast and efficient 
Mellin technique for incorporating NLO pp scattering cross sections 
in the analysis is of course not restricted to RHIC, but could equally
find important applications at the LHC.

%%%%%%%%%%%%%%%%
%\acknowledgments
%%%%%%%%%%%%%%%%
We thank E.C.\ Aschenauer, A.\ Bazilevsky, A.\ Deshpande, R.\ Fatemi,           
S.\ Kuhn, and B.\ Surrow for communications.                                    
W.V.\ thanks the U.S.\ Department of Energy (contract no.\                      
DE-AC02-98CH10886).                                                             
This work was partially supported by CONICET, ANPCyT and UBACyT.

%%%%%%%%%%%%%%%%%%%%%%%%%%%


\begin{thebibliography}{99}
\bibitem{ref:footnote1} We add statistical and systematical errors in quadrature
as full error correlation matrices are not always available.
\bibitem{ref:webpage} A {\sc Fortran} code, list of fitted parameters, and
additional material can be found at {\tt ribf.riken.jp/}$\sim${\tt marco/DSSV}.
\end{thebibliography}
\end{document}